**The Everett Interpretation: Structure**

Simon Saunders

What does quantum mechanics tell us, if taken realistically, as a fundamental theory, which applies to everything? After a century of debate, it seems we are no nearer to an answer to this question. But in an important respect we know better what is in contention: it is the *Everett interpretation* of quantum mechanics, which is also known as the *many-worlds interpretation*. First advocated by Hugh Everett III in 1957, it is the only realist interpretation of quantum mechanics still standing. On every other approach the answer is that quantum mechanics tells us nothing when applied to everything because it makes *no sense* taken in this way; the theory must be changed. In the words of John Bell: "Either the wavefunction, as given by the Schrödinger equation, is not everything, or it is not right,"[1] seeking to reduce the options to either modifying the Schrödinger equation, or supplementing it with hidden variables. But "everything" for Bell meant "everything in the known universe"; ignored was the alternative that the wavefunction (as given by the Schrödinger equation) describes more than the known universe – that it describes a quantum mechanical multiverse, a superposition of worlds, of which ours is only one.

The idea is fantastical, but quantum mechanics is a theory like no other, a revolution still in the making after all these years. The worlds are derived from the unitary formalism; they are not put in by hand. Measurement interactions, we know, lead to macroscopic superpositions – if the unitary dynamics operates untrammelled -- where each state in the superposition contains a record of a sequence of events as if obtained by the measurement postulates, as Everett gave a simple model to show. The apparent conflict between quantum theory and locality, as codified in Bell's theorem, is removed: the interpretation extends without modification to relativistic quantum theory. It really is *interpretation* of the equations: it neither modifies nor supplements them, save that the Schrödinger equation is taken to be universal. It offers a radical and novel understanding of how determinism can be reconciled with indeterminism, solving puzzles in the philosophy of probability that have bedeviled the subject for decades. It provides the basis for quantum cosmology, free of the measurement problem.

These claims, of course, are all controversial. In a certain sense, if these arguments all stand up to scrutiny, the case for the Everett interpretation becomes *overwhelming*. No wonder they are strongly contested: there is simply too much at stake.

Much recent literature has been on the probability interpretation,[2] but there is another aspect to the development of the Everett interpretation in the last three decades that is just as important: in terms of *decoherence theory*. This frees the interpretation from dependence on the notion of measurement, and enhances the argument about records. It tells us how the worlds are composed. It allows the separation of the probability interpretation from the question of the structure of the state. It provides an Everettian "tapestry of events," made out in entirely categorical properties and relations, that now include phase relations and amplitudes, on which the probability interpretation is to be based.

This chapter is on this structural interpretation of the wave function, rather than the probability interpretation, which is the subject of a companion paper (Saunders 2021). In particular, it is on the structure of the wave function as made out in terms of the quantum histories formalism. But it is also on Everett's writings, and especially his "automaton" argument, as published in 1957. I shall start with this, for it makes clearer the role of decoherence theory in going beyond Everett's writings (although as we shall see, there are hints of it in his "long" dissertation, eventually published in 1973 as "The theory of the universal wave function"[3]). Everett's much more widely read doctoral thesis was one fifth the length, published under the title "'Relative State' formulation of quantum mechanics." (Everett, 1957). It was redacted from the "long" dissertation at the insistence of his supervisor John Wheeler, who had long advocated Bohr's philosophy. It contained the argument from records, and thanks to his "Note added in proof" it conveyed the overall idea, but much was omitted. Everett was never to write on quantum mechanics again.

### 1 Everett's Automaton Argument

The strange gap between the determinism of the Schrödinger equation and the indeterminism evident at the observational level is bridged by the so-called *measurement postulates*. The first is given by the rule:

> **Born rule:** on measurement of $Q = \sum_k \lambda_k P_k$ on a system *S* prepared in the state $|\phi\rangle$, the outcome $\lambda_k$ is obtained with probability $\frac{\langle\phi|P_k\phi\rangle}{\langle\phi|\phi\rangle}$.

(The $P_k$s are projection operators onto eigenstates of *Q* with eigenvalues $\lambda_k$ -- this extends naturally to operators with continuous spectra. I will not always respect the distinctions between vectors, wavefunctions, and rays, and use the term "state" for all three. By the "amplitude" of a state, I mean its norm $\sqrt{\langle\phi|\phi\rangle}$.)

The Born rule has the flavor of a "correspondence rule" in logical empiricist philosophy of science, or perhaps an "operational definition": a rule that links theoretical concepts to observable ones (or theoretical terms to observation sentences). It provides the minimal interpretation necessary to submit quantum mechanics to test – provided, of course, recipes could be given for building appropriate measurement and state-preparation devices. The latter required purely classical concepts, according to one influential interpretation (Bohr's). That posed an obstacle to taking quantum mechanics as a fundamental theory: how then could it be based purely on classical concepts? (Bohr thought of quantum mechanics as a "generalization" of classical mechanics, rather than a theory distinct from it;[4] here Wheeler departed from Bohr, as he sought to quantize gravity, governing the macroscopic.)

The Born rule is the first and most important of the measurement postulates, but it says nothing about what happens to the state on measurement. It is usually supplemented by the:

> **Projection postulate**: for a repeatable experiment on $S$ prepared in the state $|\phi\rangle$, if the outcome on measurement is $\lambda_k$, then $S$ is left in the state $|\phi_k\rangle = P_k|\phi\rangle$.

The projection postulate is sometimes stated without the restriction to repeatability, but then it is very often false. Very often, in practice, the state of the system is changed in uncontrollable ways on measurement, or the system is even completely obliterated (so there is no state). But given that the macroscopic outcome is indeterministic, it is reasonable to suppose that the state of the measured system where it does still exist has changed indeterministically, and if it makes sense to assign a quantum state to the measurement apparatus as well, it too must have been subject to indeterministic change. How is all this indeterminism at the level of the state consistent with the Schrödinger equation?

This is the infamous *measurement problem* of quantum mechanics, in what is probably its simplest guise. There are the two realist solutions already noted: add additional variables, or modify the Schrödinger equation. Everett offered a third and remarkable alternative: the *superposition* of all the indeterministic changes evolves deterministically, in accordance with the Schrödinger equation.

Everett's strategy to establish this conclusion was to show that a superposition of *records* of indeterminism could be obtained in this way, satisfying the Schrödinger equation – and specifically, records as could be encoded in a mechanical model. "The observer" was to treated as a mechanical system interacting with the measured system, with the unitary equations applied to them both together. The answer to the question "What is observed?" is to be read off from this dynamical model, in terms of what can be laid down in records or in memory. This solved the problem, posed by Bohr, of how a system could be treated as both physically closed and yet under experimental control. Bohr had argued that physical closure renders external observation impossible (unlike in classical physics, where the interactions needed for observation to be possible could be made arbitrarily small).[6] Everett's answer was to model the observer as *within* the closed system.

A great deal hangs on this argument. Is it acceptable to read off what is observed, from a direct dynamical model of observation, modeling the observer alongside the system under observation? Is Everett's approach in this respect a heightened realism, an extreme form of physicalism? No: as scientific method, it has impeccable credentials. It was important in the early modern period in establishing the Copernican system (a comparison Everett himself made), namely, in the analysis of what would be perceived, according to the theory, were the Earth in motion about the sun (showing there would be no great wind, no deviations in falling bodies). It was essential in the discovery of symmetries (think of Galileo's ship, and Faraday's cage). It was essential to Isaac Newton's method in the *Principia*. It played an essential role in Einstein's analysis of length contraction and time dilation in special relativity. Even Bohr extolled the general principle: "it is the theory that tells us what is observable."[7] The method is part and parcel of all the great discoveries in fundamental physics – save one. So is the practice of studying those theories at maximum strength, with the fundamental equations of each theory taken to have maximal scope – save one. Quantum mechanics is the one exception. Von Neumann had made the first important step, of modelling the observer in quantum mechanics, in his *Mathematical Foundations of Quantum Mechanics*, published (in German) in 1932, but he was not yet ready to study the consequences of taking the Schrödinger equation to hold unrestrictedly. That honour fell to Everett.

Consider, for the sake of definiteness, a measurement of the z-component of spin of an atom of silver, as in the Stern-Gerlach experiment. The interaction Hamiltonian couples the magnetic moment of the atom in passaging a magnetic field with a certain symmetry, in such a way as to correlate the momentum of the atom of silver with the state of its component of spin, in that direction, either "+" or "-". The result is that the atom drifts in one of two opposite directions. A subsequent measurement of position is then directly correlated with the state of this component of spin. Depending on how the position measurement is performed (it may be made with wide latitude), the measurement may be repeatable. Suppose that it is. Then schematically, if the apparatus is to function as intended, when the initial state of the atom is $|\phi_+\rangle$ and the apparatus is in its "ready" state $|0\rangle$, it should be driven by the unitary dynamics governing the measurement, denote $U_m$, to display the outcome "+"; and when the initial state of the atom is $|\phi_-\rangle$ and the apparatus is in the ready state $|0\rangle$, it should be driven by $U_m$ to display the outcome "-"; and in either case, for repeatability, the spin state of the silver atom should be unchanged. That is, $U_m$ should satisfy the protocols:

$$|\phi_+\rangle \otimes |0\rangle \xrightarrow{U_m} |\phi_+\rangle \otimes |+\rangle \qquad (1a)$$

$$|\phi_-\rangle \otimes |0\rangle \xrightarrow{U_m} |\phi_-\rangle \otimes |-\rangle. \qquad (1b)$$

But it then follows, for any unitary dynamics like this, that for an initial state

$$|\psi\rangle = c_+|\phi_+\rangle + c_-|\phi_-\rangle \qquad (2)$$

where $c_\pm$ are complex numbers:

$$|\psi\rangle \otimes |0\rangle \xrightarrow[U_m]{} c_+|\phi_+\rangle \otimes |+\rangle + c_-|\phi_-\rangle \otimes |-\rangle. \qquad (3)$$

What could the RHS mean? Von Neumann's answer was that something very like correspondence rules were needed (although he did not use that terminology); for, he argued, mathematical expressions on their own, governing values of quantities, *never* amounted to a statement about what is observed. Needed was a further link to "experience": to a statement about what would be *perceived* or *observed*, or *experienced*; and this notwithstanding that mental events always had correlates in the physical (his famous "thesis of psychophysical parallelism", one of the core tenets of realism). According to von Neumann, rules like this were needed as much for Eq. (1a) and (1b), delivering probability one for seeing spin-up, given Eq. (1a), and probability one for seeing spin-down, given Eq. (1b), as they were needed given the superposition Eq. (3), when the probabilities are different from one. The rules were the measurement postulates (or what von Neumann called "Process 1"). Only in this way, said von Neumann, was the dynamical model including the observer "non-vacuous."

Everett, who studied von Neumann's book assiduously (it first appeared in English translation in 1955), took the thesis of psychophysical parallelism rather more to heart than did its author. If you want to know about the mental, he reasoned, sufficient to make sense of measurements, then model perception and memory explicitly. Suppose the apparatus to be a simple mechanical device, able to interact with the measured system in the sense of perception, but also able to store records of such perceptions in memory. Let the ready state of the apparatus with no records in memory be $|0; ...\rangle$; in place of Eq. (1a) and Eq. (1b), require that in addition require the positive spin outcome be recorded in memory as "+," and the negative as "–," and that the device resets to "0". That is, suppose that we can build a mechanical device so that under the Schrödinger equation it satisfies the new protocols:

$$|\phi_+\rangle \otimes |0; ...\rangle \xrightarrow[U_m]{} |\phi_+\rangle \otimes |+; ...\rangle \xrightarrow[U_0]{} |\phi_+\rangle |0; +, ...\rangle \qquad (4a)$$

$$|\phi_-\rangle \otimes |0; ...\rangle \xrightarrow[U_m]{} |\phi_-\rangle \otimes |-; ...\rangle \xrightarrow[U_0]{} |\phi_-\rangle |0; -, ...\rangle \qquad (4b)$$

where $U_m$ is the measurement process as before, and $U_0$ further records the outcome and resets the apparatus. "Experience" is now to be read off from what is registered and what is laid down in memory.

What happens now, under these new protocols, when the initial state is the superposition $|\psi\rangle$? The answer, from linearity, is:

$$|\psi\rangle \otimes |0; ...\rangle \xrightarrow[U_m]{} c_+|\phi_+\rangle \otimes |+; ...\rangle + c_-|\phi_-\rangle \otimes |-; ...\rangle$$

$$\xrightarrow[U_0]{} c_+|\phi_+\rangle \otimes |0; +\cdots\rangle + c_-|\phi_-\rangle \otimes |0; -\cdots\rangle.$$

The final state is a superposition of a record of positive *z*-component of spin, with a record of negative spin. On repeating the experiment (on the same microscopic system), we obtain:

$$c_+|\phi_+\rangle \otimes |0;+\cdots\rangle + c_-|\phi_-\rangle \otimes |0;-\cdots\rangle$$

$$\xrightarrow[U_m]{} c_+|\phi_+\rangle \otimes |+;+\cdots\rangle + c_-|\phi_-\rangle \otimes |-;-\cdots\rangle$$

$$\xrightarrow[U_0]{} c_+|\phi_+\rangle \otimes |0;++\cdots\rangle + c_-|\phi_-\rangle \otimes |0;--\cdots\rangle.$$

The final state is a superposition of the record of a "+" outcome followed by a second "+" outcome, with the record of a "-" outcome followed by a second "-" outcome – *where each of the latter is just what would have been obtained, on von Neumann's terms, by employing his Process 1* (essentially the projection postulate, extended so as to apply to the state of the measurement device).

What does it mean to have records of measurements *in a superposition* – for there to be two states, each describing a record of measurement, in a superposition? In itself the concept is hardly unfamiliar, in that every student of quantum mechanics is used to the idea of superpositions of contradictory properties at the microscopic level. It isn't even particularly mysterious when divorced from the measurement problem: there are superpositions of light signals and radio programs and TV channels in the electromagnetic field as well – even, or especially, when considered purely classically. This is not a problem. (But it can be made to *seem* mysterious – by insisting not that there is a superposition of radio programs, but that there is a radio program in a superposition – a point we shall come back to.)

Nor are we unused to multiple realities somehow in relation to each other. The world as I write these words is as real as can be, but for you it is some time ago – and your world, as you read these words, is as real as can be for you too, although it is far in the future for me. We have learned, although it is still a matter of philosophical controversy, how to understand these worlds as both existing, as being worlds at different times; Everett invites us to understand two different outcomes as both existing, as being worlds that are orthogonal.[8]

But if in this way we may understand the idea of distinct processes taking place in a superposition, each as if the projection postulate had been applied, *both processes happen with certainty*. Where, in all this, is probability?

In the case of Eq. (5), the Born rule adds to the projection postulate (according to which just one of Eq. (4a) and Eq. (4b) is realized) the fact that the *probabilities* of the outcomes are $p = \frac{|c_+|^2}{|c_+|^2+|c_-|^2}$ and $1-p$ respectively. In the Everett interpretation, it follows from the Schrödinger equation that the *amplitudes* or the superposed outcomes are $\sqrt{p}, \sqrt{(1-p)}$ respectively. In neither case do we as yet have a connection with any observable quantity. To do this, what is needed, of course, are multiple experiments: a large number $N$ of silver atoms all prepared in the same state Eq. (2), and independently measured in accordance with Eq. (4) (whether at the same time or at different times). The result will be a superposition of $2^N$ states, each a record of a unique sequence of measurement outcomes, each the same as that which would have been obtained by veridical observation, had that sequence resulted by chance, using the measurement postulates.

Everett called them *branches*. It is not too hard to see that the connection between amplitude and Born-rule probability is retained for multiple experiments. The amplitude of each branch, at the end of N experiments, as determined by the unitary evolution alone (together with the initial state), equals the square root of the Born-rule probability for that sequence of outcomes (just multiply together the probabilities for the results taken sequentially). Now consider the superposition of all those branches with the same relative frequency for the "+" outcome; not quite so obviously, the amplitude of this superposition is highly sensitive to the discrepancy, if any, between that relative frequency and the Born rule quantity for the "+" outcome, the quantity $p$. Let the discrepancy be $\varepsilon$; then the amplitude falls off exponentially as $\exp -N\varepsilon^2/\kappa$, where $\kappa = 4p(1-p)$ and $N$ is, as before, the number of trials.[9] It is the first of a number of quantum Bernoulli theorems, the quantum analogues of the laws of large numbers: the amplitudes of branches with the "wrong" relative frequencies fall off exponentially quickly in the number of trials, in comparison with the amplitudes of Born-rule compliant branches.

The squared amplitudes, in these essential respects, behave just like probabilities. Why the square? Everett had an answer to that question too. Let $\mu$ be a probability measure over orthogonal branches. Then it should satisfy additivity:

$$|\psi\rangle = \sum |\phi\rangle \Rightarrow \mu[|\psi\rangle] = \sum \mu[|\phi\rangle] \qquad (5)$$

But then, if it is a function of the branch amplitude, it must be the square. (Let $\mu(\sqrt{x}) = f(x)$. Then from Eq. (5), $f(\sum|c_k|^2) = \sum f(|c_k|^2)$, so $f$ is linear in $x$, $f = kx$ for some constant $k$. So $\mu(x) = f(x^2) = kx^2$. Here, $k$ is fixed by normalization; Everett also showed that the phase is irrelevant.)

If now we may interpret the squared amplitudes of branches produced by measurements as the physical correlates of the concept of physical probability, we will have explained the Born rule. Can we? That question we are postponing to a companion paper (Saunders 2021). In the rest of this, we consider rather how Everett's ideas relate to decoherence theory, and what difference that theory makes. But his needs some more history.

## 2 Realism about Measurements

The arguments thus summarized were all in Everett's "'Relative state' formulation of quantum mechanics." What of the relative state? Given an entanglement of the form (3), there is no choice of basis, respecting the tensor-product structure between the silver atom and the apparatus, in which the state is a product state; there is no way of attributing a unique pure state to the silver atom, or to the measurement apparatus. But from the total state and a pure state of the apparatus, a unique pure state of the silver atom can be defined. This is its relative state.

On a more deflationary way of putting it: given an entanglement, there are many correlations between states of subsystems. "Relative state" is useful terminology, and reminds us that the structure of the quantum state is relational. There is however a connection with the extended projection postulate. Let the relative state of $|\varphi\rangle$ in an entanglement $|\Psi\rangle$ be $|\psi\rangle$; then up to normalization, $|\varphi\rangle \otimes |\psi\rangle = P_{|\varphi\rangle} \otimes I \, |\Psi\rangle$. The collapse of the wave-function, in terms of the extended projection postulate, is relativization of state, relativized to a state of the apparatus. (As Everett put it: "the discontinuous 'jump' into an eigenstate is only a relative proposition, dependent upon the mode of decomposition of the total wave-function into the superposition, and relative to a particularly chosen apparatus-coordinate value" (Everett, 1957, p. 457).) It naturally generalizes to the relative state of a range of values of dynamical variables (and not just eigenvalues): replace $P_{|\varphi\rangle}$ by a projection $P_\Delta$ onto values of variables in some set $\Delta$ (a coarse-graining of the parameter space).

But this came later; well into the 1980s, the focus was on simple bipartite systems and the exact definition of a unique basis, in terms of which to decompose the total state, among them that given by the biorthogonal decomposition (or "Schmidt decomposition"). This was a harbinger of other ways of reading Everett's "Relative states" paper. For any entangled state $|\Psi\rangle$ of two subsystems, there exist orthonormal bases $\{|\varphi_k\rangle\}$ and $\{|\eta_k\rangle\}$ such that $|\Psi\rangle = \sum_k c_k |\varphi_k\rangle \otimes |\eta_k\rangle$. If $|c_k| \neq |c_j|$ for $k \neq j$, the bases are unique. (Equivalently, diagonalize the reduced density matrices of the two subsystems.) Dieter Zeh's early work on decoherence theory made use of biorthogonal decomposition (see e.g. Zeh, 1973), what Everett had called the "canonical representation" (Everett, 1973, p. 47). This was also the key to the "modal interpretation" (Dieks and Vermaas, 1998), for which the failure of uniqueness eventually proved terminal. Zeh's most important contribution to decoherence theory, in collaboration with Ehrich Joos in 1985, was based on rather different ideas.

There were other distractions. Everett spoke of memories and records: what were they records of? Was the approach committed to a model of consciousness,[10] and were there only "appearances" of outcomes, rather than the outcomes themselves? Everett also spoke of a "memory trajectory of an observer" as being a "branching tree," suggesting there is only one observer in a superposition, – a question that Everett in the long dissertation called a "language difficulty." Perhaps the theory isn't committed to a genuine multiplicity after all?[11] In the two-slit experiment, is there one particle at both slits, or are there two particles? Everett also claimed the branches were non-interacting; is that true? And most concerning of all: the branches, the basis states entering into the superposition, were defined by the measurement interaction (in effect the protocols (1), (3)); but the approach is supposed to be realist, with measurements playing no special role. How was this basis to be defined without them?[12]

Improving on Everett's ideas in any of these ways seemed to require new assumptions, new postulates, new physics; but any step of that kind compromised the chief selling point of his approach – that it is quantum mechanics and nothing else. The problem of basis, the "preferred basis problem," was the most serious. If experiments play no special role (and there is no Born rule to specify the basis), what basis is to be used to define the branches? And relatedly, when exactly does branching occur?

Bryce DeWitt, the first to take Everett's ideas seriously (the terminology "many worlds" is due to him), wrote on them in a number of articles (anthologized in DeWitt and Graham (1973), but mostly avoided the question of basis. I have found only one comment that directly addressed this question:

> The student should perhaps be reminded again at this point that reality is not described by the state vector alone, but by the state vector plus a set of dynamical operator variables satisfying definite dynamical equations. Decompositions of the form [Eq. (2), (4)] are not to be regarded as meaningful if they are merely abstract mathematical exercises in Hilbert space. Indeed such mathematical decompositions can be performed in an infinity of ways. Only those decompositions are meaningful which reflect the behavior of a concrete dynamical system.[13]

But the only mentioned systems were experiments, the only dynamical equations were for kinds of measurement interactions. "The many-worlds theory," according to DeWitt, rested on two postulates, both of which he attributed to Everett: the "postulate of mathematical content," concerning operator algebras and Hilbert-space theory, and the "postulate of complexity," namely, that "the world is decomposable into systems and apparata."[14] Here talk of "complexity" was a fig leaf: since when did a fundamental theory *postulate* the existence of measurement devices? Everett had made no such suggestion. What of a world without any people, without any devices?

We find in Everett's long dissertation a rather different set of ideas. He spent time on the definition of the entropy function in classical and quantum statistical mechanics, and on the concept of thermodynamic equilibrium. He pointed out that the unitary equations for many-particle systems did not imply that particles diffuse formlessly: electrons and protons in a box are not uniformly diffused, rather, they are diffused as atoms of hydrogen, and so on for more structured molecules and composites. He reminded us that the diffusion of the center of mass ('centroid') wave-function for a large numbers of particles in a bound state is extremely slow, for even the smallest visible specks of matter. Everett's proposal was to use, as basis states, wave-functions for centres of mass, as functions on three-dimensional space, well-localized in position and momentum.

What had always blocked this line of reasoning as an account of how macroscopic physics emerges from the unitary equations of quantum mechanics is that states in general do *not* have this form (and in particular: states following a quantum measurement do not have this form). But on Everett's approach the development of macroscopic superpositions is a feature, not a bug:

> The general state of a system of macroscopic objects does not, however, ascribe any nearly definite positions and momenta to the individual bodies. Nevertheless, any general state can at any instant be analyzed into a superposition of states each of which does represent the bodies with fairly well defined positions and momenta.

Everett concluded with a reference to von Neumann's construction of projections onto approximately well-localised regions of the one-particle phase space, what von Neumann had called "elementary building blocks of the macroscopic description of the world."[15]

So can the preferred basis be simply *stipulated*, chosen so that each basis state describes the macroscopic in recognizably classical terms? It is sometimes said that if there is to be a preferred basis, it must be *postulated* – written into the axioms of quantum theory – for the Everett interpretation to be well-defined. But neither stipulation nor postulation is needed: the state can be expanded in any basis, and using another, we do not obtain a new and different reality. It is the same reality, only divided up in a new way. Divide it then in the way that makes perspicuous its structure, perhaps one among several.

Are there no constraints? For example, is it true that the branches will be non-interacting? Everett had said this followed from linearity of the Schrödinger equation alone – so, presumably, whatever basis is used – but here he was less sure-footed. It is true that from linearity, each state evolves as if the other is not there; given unitarity as well, if two such states are orthogonal, they remain orthogonal as they evolve in time. But that does not mean that on *subsequent* branching, the states thus produced will not interfere with each other. Take, for example, the two-slit experiment. On passage through the slits, the photon is in a superposition of two orthogonal states, each originating from one of the slits; they remain orthogonal under any unitary transformation, including their evolution to the screen. But there is interference at the screen: represent each as a superposition of spatially localized states, at the screen, and the latter are *not* all orthogonal to one another.

However, what the two-slit experiment also demonstrates is that overlap in configuration space is necessary for interference. This suggests that states describing large numbers of molecules all well-localized in space with well-defined velocities, even at the microscopic level, can no longer interfere. If the difficulty in getting states like this to overlap in configuration space needed spelling out, Everett could have cited David Bohm, who had written on it both in his then recent book *Quantum Theory*, and in the two-part paper on hidden variables that quickly followed.[16]

It is sometimes said that decoherence theory is needed to show that Everett's branch states do not interfere with each other, and it surely offers quantitative control; but the greater importance of decoherence theory lies elsewhere.

What of the comparison of the state to a tree-like structure? This was forced by the protocol (3), where the branching is defined by a measuring interaction for given initial state, and it fits with the analysis of chance: chance events are branching events. The recombination of branches has no such interpretation. But the same reasoning would seem to apply to states of macroscopic bodies differently localized in phase space: we do not expect a superposition of such states to unitarily evolve into a single localized state, for that would seem to require that they be finely synchronized (we shall come back to this shortly).

How *do* states well-localized in position and momentum unitarily evolve? When the mass is not too small or the time-scales are too large, the answer is: classically. To continue Everett's argument, it is not just that states of large numbers of particles well-localized in phase space do not interfere, and not only that at each instant in time they are recognizably interpretable in classical terms; it is that they *behave* classically:

> Each of these states then propagates approximately according to classical laws, so that the general state can be viewed as a superposition of quasi-classical states propagating according to nearly classical trajectories. In other words, if the masses are large or the time short, there will be strong correlations between the initial (approximate) positions and momenta and those at a later time, with the dependence being given approximately by classical mechanics.

A macroscopic body is then a propagating quasiclassical state, approximately obeying a classical Hamiltonian equation – it is dynamically defined. Notice now how the "language difficulty" is handled. A quasiclassical state propagating as such is a thing, so when a superposition develops, do not say "a thing is in a superposition," say rather "there is a superposition of things." (It is not a beam of light in a superposition of two directions, it is two beams in a superposition.) There is the sideways view of the state at each time, as a temporal sequence of superpositions, and there is the vertical view, of a superposition of temporal sequences, of "memory trajectories" or "propagating quasiclassical states" – each in accordance with classical equations.

Applied to the automaton, made up of mechanical parts (the servo-mechanism), Everett's model of the observer, it completed his argument:

> Since large scale objects obeying classical laws have a place in our theory of pure wave mechanics, we have justified the introduction of models for observers consisting of classically describable, automatically functioning machinery, and the treatment of observation is non-vacuous.[17]

The dynamical circle is closed, linking micro to macro, where the latter obeys classical equations, as follow from the Schrödinger equation, ensuring that the protocol (3) is satisfied. The measurement protocols are not merely *stipulated*, they have to be physically instantiated

But although the circle is closed, it still depends on measurement processes; without them, what is the superposition, and what are the branches? DeWitt had a reason for his "postulate of complexity"; how to dispense with measurement interactions? But now there is an obvious candidate for an answer, prefigured by Everett: these propagating quasiclassical states have only *nearly* classical trajectories, and where that approximation is not satisfied, there is branching.

Everett did not take this further step; nor did he offer an argument for his claim of approximate classicality. But he could surely have provided one, drawing, if he wished, from Bohm's book, which included a chapter on the WKB approximation. (Everett cited this book alongside von Neumann's as his primary influences.) Here is a sketch based on Ehrenfest's theorem in the simple case of a single massive particle with position operator $x$. The theorem states that in the state $|\psi\rangle$ and for the potential function $V(x,t)$ (using the notation $\langle\psi|x|\psi\rangle = \langle x\rangle_\psi$):

$$m\frac{d^2\langle x\rangle_\psi}{dt^2} = -\langle \nabla V(x,t)\rangle_\psi.$$

It does not tell us that a particle behaves classically; failing an interpretation of the expectation value $\langle x\rangle_\psi$ for a single system, in an arbitrary state, it tells us nothing at all. But in the special case where $|\psi\rangle$ is well-localized in position and velocity, and the gradient of the potential is approximately constant over such regions and slowly varying in time, we can replace the RHS by the gradient of the potential as a function of $\langle x\rangle_\psi$, to obtain:

$$m\frac{d^2\langle x\rangle_\psi}{dt^2} = -\nabla V(\langle x\rangle_\psi, t)$$

and whatever exactly $\langle x\rangle_\psi$ means, it obeys the same equation as does the position of a classical particle; so long as the approximations hold, $\langle x\rangle_\psi$ behaves just like the classical particle position behaves. (This argument extends to systems of many weakly-interacting bodies, each localized in position and

momentum.[18]) And where it fails to hold, then there is branching of such trajectories, and in the branching, the quantum jump, and the appearance of randomness.

The nearly satisfied equations and the propagating quasiclassical states go together; neither is defined without the other. It is these patterns in the wave-function that define the preferred basis in the Everett interpretation. They are dynamical patterns. They have to be derived from the unitary formalism, they cannot be "stipulated." In this respect the rules that defined the branches in the case of the measurement of spin – the measurement protocols – are a special case. They are given substance by the design and fabrication of a real physical system that instantiates them, according to the Schrödinger equation.

Worlds exist insofar as rule-based branches exist. *That* such branches evolve in accordance with those rules is basis-independent; *use* of a basis that assigns orthogonal states to each branch at each time is a matter of convenience. There is a parallel with the choice of coordinates in general relativity: one choice may be better adapted to the matter distribution, and describe it in a more perspicuous form, than another.

Might there be essentially just the *one* set of rules of this kind, one pattern, namely of states well-localized in phase space, satisfying classical Hamiltonian equations? With these as the building blocks, we can imagine recovering all of macroscopic physics without having to revisit its quantum origins. But that would already have seemed fanciful in Everett's time. Take for example the property of rigidity; no choice of a potential function for a system of molecules can account for it in classical dynamical terms.

With more urgency, then, what are these rules and states? To what approximations do they hold, for what initial states and quantum Hamiltonians, when does branching occur, and of what are branches made?

## 3 Quantum Histories and Quasiclassical Domains

The list of classical or quasiclassical equations that have been derived from quantum mechanics is lengthy. Examples include Boltzmann's equation, the Focker-Plank equation, the Langevin equation, the Navier-Stokes equation, the Lindblad equation, the Joos-Zeh equation (quantum Brownian motion), the Caldera-Leggett model, and for sufficiently well-behaved potentials and short enough times, classical Hamiltonian equations. Of course, they were derived in a number of different ways; nevertheless, they are subject to theoretical checks, using the unitary formalism assisted as needed by the measurement postulates or their generalizations.

These can all, loosely, be called the business of decoherence theory, mostly developed independent of Everett's ideas (for example, in quantum optics in the 1960s, and in open quantum systems theory in the 1970s). But decoherence theory, on inspection, is itself a mass of different models, for different kinds of dynamical variables, in different coupling regimes and environments.[20] Some of them, spin-foam models for example, have little to do with ordinary matter. And those that do, even if embedded in or derived from non-relativistic quantum mechanics, do not in themselves speak for the Everett interpretation. The crucial question is whether the values of these quasiclassical variables, as they vary in time, obeying these phenomenological equations, can be derived from the unitary formalism in the way Everett had suggested: as superpositions of quasiclassical states where each of the latter propagates along definite trajectories, with branching quantified in terms of quantum dissipation and noise.

A proper answer to this question would require an investigation in each case taken separately, but they share much in common. They are mostly equations for many-particle systems. They are all derived by coarse-graining: the integrating out of some degrees of freedom, the definition of new "effective" degrees of freedom, as coarse-grained values of the old, including coarse-graining in time and the separation of "slow" from "fast" variables. The most versatile and widely used technique is probably the path-integral method, but there is a simple Hilbert-space framework as well: the *quantum histories formalism*. At low energies the two are translatable into each other.[21] From the quantum histories formalism the connection to Everett is direct.

Let us begin with a single time $t_k$. Let $\alpha_k$ (for fixed k) range over coarse-grained cells of some parameter space $\mathcal{M}$ at time $t_k$ (phase space, for example). Let $P_{\alpha_k}$ be the associated projectors, so that for $\alpha_k \neq \alpha_k'$, $P_{\alpha_k}$ and $P_{\alpha_k'}$ are orthogonal, with $\sum_{\alpha_k} P_{\alpha_k} = I$. When $\mathcal{M}$ is a parameter space for commuting variables, the Cartesian product of their spectra, the associated families of projections are the basic tools of spectral theory. When $\mathcal{M}$ is phase space, a more complicated construction is needed, as given by von Neumann's "elementary building blocks".

It is obvious how to coarse-grain: pass to sums of projection operators, projecting onto unions of subspaces. Let $\{\beta_k\}$ be a coarse-graining of $\{\alpha_k\}$ (so each cell $\beta_k$ in the parameter space is the union of some of the $\alpha_k$s, denote $\alpha_k \subset \beta_k$ ); then the projection operator corresponding to $\beta_k$ is:

$$P_{\beta_k} = \sum_{\alpha_k;\, \alpha_k \subset \beta_k} P_{\alpha_k}.$$

Sums of commuting projectors correspond to the union of the $\alpha_k$s , and products to intersections.

All of this is entirely familiar. What is distinctive of the quantum histories formalism is to go over to the Heisenberg picture, and put what has so far been a purely notional time parameter to work. For each time $t_k$, define

$$P_{\alpha_k}(t_k) \stackrel{\text{def}}{=} U(-t_k) P_{\alpha_k} U(t_k)$$

where $U(t_k) = e^{-iHt_k}$ and $H$ is the quantum Hamiltonian. Consider now a sequence of times $t_N > t_{N-1} > \cdots > t_1$, and let $\alpha = \langle \alpha_N, \alpha_{N-1}, \ldots, \alpha_1 \rangle$ represent a corresponding sequence of parameters. Define the associated "chain" or "class" operators $C_\alpha = P_{\alpha_N}(t_N) \ldots P_{\alpha_1}(t_1)$, in general the product of non-commuting operators. Let the initial state $t_k = 0$ be $|\psi\rangle$; then $C_\alpha|\psi\rangle$ is obtained by: unitarily propagate $|\psi\rangle$ to $t_1$, project out the state $P_{\alpha_1}U(t_1)|\psi\rangle$, unitarily propagate to $t_2$, project out the state $P_{\alpha_2}U(t_2 - t_1)P_{\alpha_1}U(t_1)|\psi\rangle$, and so on – obtaining precisely the same state, at each time, as were a sequence of measurements performed, up to that time, yielding the outcomes $\alpha_1, \alpha_2, \ldots$, and applying the extended projection postulate (whether sequentially or just at the end). The Schrödinger-picture end-state obtained in this way has amplitude equal to the square root of the product of all the Born rule probabilities for that sequence of outcomes. Unitarily evolve it back to $t_0$ and we obtain the Heisenberg-picture state $C_\alpha|\psi\rangle$, representing that history.

Each history $a$, as given by the sequence of parameters $\langle \alpha_N, \alpha_{N-1}, \ldots, \alpha_1 \rangle$, may seem to have nothing to do with any unitary equations (they may appear quite random, for example). But the *superposition* of all these propagating sequences, up to any time, is exactly the same as the unitary evolution of the total state to that time. That is:

$$\sum_{\langle \alpha_N, \alpha_{N-1}, \ldots, \alpha_1 \rangle} P_{\alpha_N}U(t_N - t_{N-1})P_{\alpha_{N-1}}U(t_{N-1} - t_{N-2}) \ldots P_{\alpha_1}(t_1)|\psi\rangle$$

$$= \sum_a U(t_N)C_a|\psi\rangle = U(t_N)\sum_a C_a|\psi\rangle = U(t_N)|\psi\rangle$$

where the last equality follows from the identity $\sum_a C_a = I$. The Schrödinger-picture states $U(t_N)C_a|\psi\rangle$ are the states of Everett's branches at $t = t_N$.

Coarse-graining of projectors, defined by their sums, automatically extends to coarse-grainings of chain operators, defined by their sums. If $\beta$ is a coarse-graining of $a$, denote $a \subset \beta$, meaning $\alpha_k \subset \beta_k$ for each $k$, then

$$C_\beta = \sum_{\alpha;\, \alpha \subset \beta} C_\alpha.$$

The operator $C_\alpha^\dagger C_\alpha$ is self-adjoint and positive: as such it defines a positive operator valued measure (POV measure), of the sort that now widely supplements the measurement postulates. (In the case of two-step histories, they were first known as "effects" (Davies, 1984).)

The Heisenberg picture is the natural one for determining the structure of the orbit of the quantum state, under time evolution, in four-dimensional terms. Relativistic quantum field theory, in both Lagrangian and axiomatic formulations, almost always uses the Heisenberg picture. We may think of the quantum state as fixed once and for all, and branch vectors $C_\alpha|\psi\rangle$ as components of this state, representing the corresponding histories. The perspective is that of four-dimensionalism, in the metaphysicians' sense, a "quantum block universe."[22]

What are all these histories, exactly? No use has as yet been made of the identification of the $\alpha$'s as coarse-grained values of position and momenta; the resolution of the identity could have been anything. But if it could be anything, the states $C_\alpha|\psi\rangle$ may not even be orthogonal, and there would be no guarantee that the interpretation of squared amplitude as probability is consistent with coarse-graining. In general, it is *not* true that

$$\left|C_\beta|\psi\rangle\right|^2 = \sum_{\alpha;\, \alpha \subset \beta} |C_\alpha|\psi\rangle|^2$$

(the so-called "sum rule").[23] But the two, very nearly, go together; when the $C_\alpha|\psi\rangle$s are orthogonal, the sum rule is satisfied (and when the sum rule is satisfied, the real part of $\langle\psi|C_{\alpha'}^\dagger C_\alpha|\psi\rangle$ for $\alpha \neq \alpha'$ vanishes). The "consistent histories interpretation" was developed in the hope that the sum rule would determine a preferred basis (the $P_{\alpha_k}$'s for each k) all on its own, and thus a well-defined probability measure over a space of histories, only one of which is realized, consistent with a 'one-world' interpretation of quantum mechanics. Just how weak a condition consistency really is was shown by Dowker and Kent (1996), who effectively put paid to that ambition.

The sum rule is recognizably Everett's additivity requirement, substituting the $C_\alpha|\psi\rangle$s for Everett's branches. (In the case of Everett's model of an automaton, if at each time a record is preserved of the sequence of outcomes prior to that time, the sum rule is automatically satisfied. Conversely, where two histories interfere, there can be no record of events in the two histories that differ.) For an example of coarse-graining, state, and Hamiltonian *violating* the sum rule, consider the two-slit experiment, and for projections at each time, coarse-grainings in space: projections onto the aperture $\Delta_1$ at $t_1$, at each of the slits $\Delta_+$ and $\Delta_-$ at $t_2$, and at a fixed region of the screen $\Delta_3$ at $t_3$. The histories $\alpha_\pm = \langle\Delta_1, \Delta_\pm, \Delta_3\rangle$ sum as they must to $\beta = \langle\Delta_1, \Delta_+ \cup \Delta_-, \Delta_3\rangle$, but not the associated probabilities – indeed $C_{\alpha_+}|\psi\rangle$ and $C_{\alpha_-}|\psi\rangle$ are not orthogonal, and interfere at the screen.

Unlike orthogonality of Schrödinger-picture states at an instant of time, the orthogonality of Heisenberg-picture states representing histories involves the initial state and the Hamiltonian. A *decoherent history space* over a sequence of times $t_N > .. > t_k > \cdots > t_1$ is a quadruple $\langle|\psi\rangle, H, \mathcal{M}, \{\alpha_k\}\rangle$ for which the states $C_\alpha|\psi\rangle$ are orthogonal. Such a space has the natural finite measure $\mu[\alpha] \stackrel{\text{def}}{=} \mu[C_\alpha|\psi\rangle] = |C_\alpha|\psi\rangle|^2$. Helping ourselves temporarily to the notion of probability, $\mu[\alpha]$ is the probability of history $\alpha$. Let $\{\gamma\}, \{\delta\}$ be coarse-grainings of $\{\alpha\}$ for state $|\psi\rangle$, and let their composition, denote $\gamma * \delta$, be the sequence $\langle\gamma_n \cap \delta_n, \ldots, \gamma_1 \cap \delta_1\rangle$.

If $\mu[\delta] \neq 0$, we may then define the conditional probability of $\gamma$ relative to $\delta$ as:

$$\mu[\gamma/\delta] = \frac{\mu[\gamma * \delta]}{\mu[\delta]}. \tag{6}$$

These conditional probabilities include retrodictive probabilities of some past event, conditional on some future event, or conditional on some future sequence of events; or of a past, present, or future sequence of events, conditional on some event or sequence of events. (The two-vector formalism is the special case of two-step histories in which the latest time projector is one-dimensional.) But equally, the formalism can be divested of this probability interpretation: these are ratios among squared amplitudes of state vectors, correlations among states, representing sequences of events, awaiting further interpretation.

Now for Everett's picture of a tree-like structure to the state. Given a decohering history space $\langle|\psi\rangle, H, \mathcal{M}, \{\alpha_k\}\rangle$, it is always possible to define a new decohering history space $\langle|\psi\rangle, H, \mathcal{M}, \{\epsilon_k\}\rangle$, where $\{\epsilon_k\}$ is a fine-graining of $\{\alpha_k\}$, with "branching structure" – in which histories only diverge to the future and never recombine.[24] (Branching structure, like decoherence, was ensured for Everett's automaton states, as defined by Eq. (3).) Formally, for any $t_k > t_j$, for any $\epsilon_k, \epsilon_j$,

$$\mu[\epsilon_j/\epsilon_k] = \frac{\left|P_{\epsilon_k}(t_k)P_{\epsilon_j}(t_j)|\psi\rangle\right|^2}{\left|P_{\epsilon_k}(t_k)|\psi\rangle\right|^2} \approx 0 \text{ or } 1. \tag{7}$$

The past of an event $\epsilon_k$ is therefore approximately unique. There is only one way, from a configuration at time $t_k$, of tracing a preceding sequence of configurations; Everett's concept of branching thus generalises. But to what is this temporal asymmetry to be traced? Evidently not to the unitary evolution, which is time-reversal invariant. We earlier saw reasons to expect branching in the case of states initially well-localized in position and momentum – that is, in the structure of the initial state. The point applies more generally, and it is the same as the explanation of the arrow of time in classical statistical mechanics: it is to be sought in the structure of the initial state.[25] Thus, if $|\psi\rangle$ for a given Hamiltonian $H$ and coarse-graining, $\{\alpha_k\}$ on $\mathcal{M}$ defines a decoherent history space, which hence has branching structure, $C_\alpha|\psi\rangle$ does not.

This point is worth perusing. What happens if we take a state like $C_\alpha|\psi\rangle$ as the initial state? Formally, as a Heisenberg-picture state, it is defined at $t_0 = 0$, like $|\psi\rangle$; if we pass to the Schrödinger picture, and unitarily evolve it forward in time, it is the state $U(t) C_\alpha|\psi\rangle$, which for $t = t_N$ is the end-state of the Everett branch for the history $\alpha$ at $t_N$. It seems we have everything that we could wish for, a single history theory with a purely unitary evolution. But no: while there is a single branch vector at $t = t_N$, there are innumerable others at earlier times – indeed a superposition of non-orthogonal branches, all with amplitudes and phases delicately adjusted as they unitarily evolve so that they all interfere with each other at $t_N$ and all save one wink out of existence. Moving forward in time, after $t_N$, it is Everettian business as usual, and orthogonal branching, and no fine-tuning (unless, of course, the history space was fine-tuned to begin with).[26]

Put now to one side the probability interpretation, and view norms and ratios in norms as mere correlations, mere relations among amplitudes and relative states. View Eqs. (6) and (7) as an extension of Everett relativization to times, and of correlations among states representing sequences of events. Understood in this way, the decoherent histories formalism provides a general language, a kind of four-dimensionalism, for describing the universal state, in which orthogonality of histories is as natural a criterion for a basis of states as is orthogonality of states in the case of a single time. As we have seen, it interestingly involves a direction in time, as determined by the initial state; what else does it involve?

The concept of *quasiclassical domain*, introduced by Murray Gell-Mann and Jim Hartle in the late 1980s,[27] is of a consistent history space for which the coarse-grained variables, the $\alpha_k$'s for each $k$, approximately satisfy some *closed* system of equations, as they vary along each history; it is a history space made up only of certain kinds of sequences. Or more precisely (since any quantum history space contains all possible histories definable as sequences of values of the coarse-grained variables), those histories that do not conform to the equation have negligible norm in comparison to those that do. (We now see that the Born rule, in a way, falls in this category too.) It is then an open question as to whether and what kinds of quasiclassical domains may be found, with what kinds of equations, coarse-grained variables, initial states, and Hamiltonians.

For a metaphysical way of putting it, a quasiclassical domain is defined when the universal state can be written as a superposition of histories almost all of which are lawlike; that almost all obey a definite rule or equation. They are histories of propagating quasiclassical states, obeying definite rules – *just as envisioned by Everett*. But now Von Neumann's "elementary building blocks" are only one example.

I gave the punch line in advance: all the important, effective, non-relativistic equations for bulk matter, gasses, and fluids have now been obtained in this way.[28] Of course several of them predate the quantum histories formalism, and implicitly or explicitly rely on the measurement postulates; but they can be cast into the quantum histories formalism, and we know how to interpret the measurement postulates in the Everett interpretation. Generically, these quasiclassical equations are only approximately satisfied: departures from them involve branching, and in some cases, as in classically chaotic systems, pervasive branching. The equations themselves may involve dissipation and noise. No wonder, then, that branching and worlds cannot be defined axiomatically; they are not defined at the microscopic level at all. They are *emergent* structures, to be extracted from the unitary equations for large numbers of particles, using methods similar to those that apply across the board in the physical sciences.[29]

The implications are far-reaching. For the first time, the Everett interpretation (and arguably quantum mechanics) is freed from any dependence on classical physics (despite the name "quasiclassical," the equations that define a quasiclassical domain could in principle be entirely foreign to classical physics, the variables completely alien). It is no longer dependent on the concept of measurement; branching, and with it the preferred basis, is emergent structure, dynamically defined. The arrow of time in thermal and decoherence terms is aligned. Everett's automaton argument is much stronger: the automaton itself need no longer be a mechanical system, but could be made of anything. More importantly, what can be recorded in its memory, corresponding to the sequence of its relative states, is not just the statistics of quantum experiments, but the observable law-like behavior of everything else that is going on in the laboratory – the entire phenomenology of materials, fluids, and gases, all in excellent agreement with experiment. In these respects, the Everett interpretation is much better than either pilot-wave theories or dynamical collapse theories. They solve the measurement problem, but rarely even try to obtain quasiclassical phenomenology ("the classical limit") in their terms. (Of course, pilot-wave theory can always help itself to results obtained under the unitary formalism alone, since it too preserves the Schrödinger equation as universal – but thereby illustrating the epiphenomenal character of the hidden variables. See also Rosaler (2015).)

## 4 Everett's "Note Added in Proof"

What to believe, in the face of all this evidence? Quantum mechanics may yet give way to a better theory, with an entirely different set of ideas. Doubts on the side of the probability interpretation may yet undermine the approach: see my companion paper. Experimental discoveries could as always change everything – of gravitationally induced state reduction, for example. Everett's place in history remains uncertain. But what if the superposition principle, and low-energy quantum mechanics, is here to stay, with no hint of any further, "hidden" variables?

Here is Everett's last word on the matter, in his "Note added in proof," added without Wheeler's permission, the one place where we know he spoke in his own voice. It is fitting to reprint it in full:

> *Note added in proof* – In reply to a preprint of this article some correspondents have raised the question of the "transition from possible to actual," arguing that in "reality" there is – as our experience testifies – no such splitting of observer states, so that only one branch can ever actually exist. Since this point may occur to other readers the following is offered in explanation.
>
> The whole issue of the transition from "possible" to "actual" is taken care of in the theory in a very simple way – there is no such transition, nor is such a transition necessary for the theory to be in accord with our experience. From the viewpoint of the theory all elements of a superposition (all "branches") are "actual," none any more "real" than the rest. It is unnecessary to suppose that all but one are somehow destroyed, since all the separate elements of a superposition individually obey the wave equation with complete indifference to the presence or absence ("actuality" or not) of any other elements. This total lack of effect of one branch on another also implies that no observer will ever be aware of any "splitting" process.

Arguments that the world picture presented by this theory is contradicted by experience, because we are unaware of any branching process, are like the criticism of the Copernican theory that the mobility of the earth as a real physical fact is incompatible with the common-sense interpretation of nature because we feel no such motion. In both cases the argument fails when it is shown that the theory itself predicts that our experience will be what it in fact is. (In the Copernican case the addition of Newtonian physics was required to be able to show that the earth's inhabitants would be unaware of any motion of the earth.)

It was Galileo, not Newton, who rebutted that criticism of the Copernican theory, on the basis of an incomplete and, at points, faulty conception of the physics. Everett's argument to show that we cannot be aware of branching ("splitting") was likewise incomplete: it does not rest on linearity alone. Everett, like Galileo, did not have all the physics needed to show that the appearances would be as they seem. But there is another comparison that is even more informative: between Everett's idea that all that there is are relative states and correlations, and the idea that all that there is are relative distances and relative velocities. The comparison is with Descartes. Both elevated a principle (the superposition principle; the principle of inertia) to universal status; both, in their different ways, had their writings suppressed. Both were transitional figures: neither was able to show, on dynamical grounds, what were the superposed worlds, what were the inertial motions. Both died young, their work unfinished. Each argued for his world-view in the same way: by a demonstration that to a mechanical being inhabiting such a universe, the world would seem exactly the same as it seems to us in the known universe – in Descartes' case, in *Le Monde*.

---

[1] Bell (1987, p. 201).

[2] For the major lines of debate, see Saunders et al. (2010), with particular emphasis on the decision-theory strategy introduced in Deutsch (1999). The *locus classicus* for the latter is Wallace (2012). See also the companion paper (Saunders 2021).

[3] In DeWitt and Graham (1973). For the story of Everett's relationship with Wheeler, see Byrne (2010).

[4] See Saunders (2005) for an extended discussion of this idea.

[6] This was the main argument for the "complementarity" (mutual exclusivity) of causal and spatiotemporal descriptions, on its first appearance (Bohr, 1928).

[7] As Everett pointedly reminded us (Everett, 1957, p. 455).

[8] See Saunders (1995, 1996, 1998) for more on the parallels between Everett's branching structure and four-dimensionalism in relativity theory.

[9] See Saunders (2010, pp. 188–189) and Wallace (2012, p. 140) for further discussion.

[10] As suggested by Albert and Loewer (1988), Lockwood (1989), Barrett (1999), and Zeh (2000).

[11] A question repeatedly raised by Jeffrey Barrett; see e.g. Barrett (2011).

[12] Additional concerns were raised about the interpretation of probability (in particular, the "branch counting rule"); see Saunders (2021).

[13] DeWitt (1971, p. 210). See also Ballentine (1973, p. 233).

[14] DeWitt (1970, p. 168).

[15] von Neumann (1955, pp. 406–409). The quoted passage is Everett (1973 p.89).

[16] Bohm (1951, ch. 6, 16, sec. 25), (1952, p. 178, fn. 18). Everett cited both these publications, albeit in other connections.

[17] Everett (1973, pp. 89–90). "Non-vacuous" was patently a jibe at von Neumann.

[18] This point requires fuller treatment that I can give it here, for it is a response to Maudlin (2019).

[20] See Stamp (2006) for a probing review.

[21] Path-integral methods of coarse-graining were introduced by Feynman and Vernon (1963), and in much of their work Gell-Mann and Hartle defined decoherence in terms of the decoherence functional and path integrals.

[22] Saunders (1995). For background in metaphysics, see e.g. Sider (2001).

[23] Due originally to Griffiths (1984).

[24] Griffiths (1993); see also Wallace (2012, pp. 93–95).

[25] See Shahvisi (this volume) and Frigg and Werndl (this volume a, b).

[26] For more on the Everett interpretation and the arrow of time, see Wallace (2012, Ch. 9).

[27] Gell-Mann and Hartle (1990, 1993).

[28] See, for example, Joos et al (2003), Schlosshauer (2007).

[29] The parallel was first drawn by Wallace (2003); it is developed at length in Wallace (2012, Ch. 1–3).